\documentclass[fp,twocolumn]{jpsj3}
\usepackage{txfonts}
\usepackage{amsmath}
\usepackage{bm}
\usepackage{color}
\usepackage{here}

\title{Material Parameters for Faster Ballistic Switching of an In-plane Magnetized Nanomagnet}

\author{Toshiki Yamaji$^1$\thanks{toshiki-yamaji@aist.go.jp} and Hiroshi Imamura$^1$\thanks{h-imamura@aist.go.jp}}
\inst{$^1$National Institute of Advanced Industrial Science and Technology (AIST), Tsukuba, Ibaraki 305-8568, Japan} 

\abst{High-speed magnetization switching of a nanomagnet is necessary for faster information processing. The ballistic switching by a pulsed magnetic filed is a promising candidate for the high-speed switching. It is known that the switching speed of the ballistic switching can be increased by increasing the magnitude of the pulsed magnetic field. However it is difficult to generate a strong and short magnetic field pulse in a small device. Here we explore another direction to achieve the high-speed ballistic switching by designing material parameters such as anisotropy constant, saturation magnetization, and the Gilbert damping constant. We perform the macrospin simulations for the ballistic switching of in-plane magnetized nano magnets with varying material parameters. The results are analyzed based on the switching dynamics on the energy density contour. We show that the pulse width required for the ballistic switching can be reduced by increasing the magnetic anisotropy constant or by decreasing the saturation magnetization. We also show that there exists an optimal value of the Gilbert damping constant that minimizes the pulse width required for the ballistic switching.
}

\begin{document}
\maketitle

\section{Introduction}
In modern information technologies huge amount of data are represented as the direction of the magnetization in a small magnet such as magnetic grains in magnetic tapes or hard disk drives. To write information on the conventional magnetic recording media an external magnetic field is applied in the opposite direction of the magnetization to switch the direction of the magnetization. During the switching the magnetization undergoes multiple precessions around the local effective field consisting of the external field, anisotropy field, and demagnetizing field. The typical switching time or write time is of the order of nanoseconds.

To meet the growing demand for fast information processing it is important to develop a faster switching scheme. The ballistic switching is a promising candidate for high-speed switching, and much effort has been devoted to developing the ballistic switching both theoretically \cite{he_theoretical_1996, sun_fast_2005, xiao_minimal_2006, nozaki_numerical_2006, nozaki_numerical_2006-1, xiao_effect_2007, horley_magnetization_2010, bazaliy_analytic_2011} and experimentally \cite{gerrits_ultrafast_2002, tudosa_ultimate_2004, schumacher_quasiballistic_2003, hiebert_fully_2003, hiebert_spatially_2003, schumacher_ballistic_nodate, kikuchi_quasi-ballistic_2014,neeraj_magnetization_2022}.
In ballistic switching a pulsed magnetic field is applied perpendicular to the easy axis to induce the large-angle precession around the external magnetic field axis. The duration of the pulse is set to a half of the precession period. After the pulse the magnetization relaxes to the equilibrium direction opposite to the initial direction. The switching speed of the ballistic switching can be increased by increasing the magnitude of the pulsed field. However, it is difficult to generate a strong and short field pulse in a small device. It is desired to find a way to speed up the ballistic switching without increasing magnetic field.

The magnetization dynamics of the ballistic switching is determined by the torques due to the external magnetic field, the uniaxial anisotropy field, the demagnetizing field, and the Gilbert damping. The torques other than the external magnetic field torque are determined by the material parameters such as the anisotropy constant, the saturation magnetization, and the Gilbert damping constant. There is room to speed up the ballistic switching by designing the appropriate material parameters.

In the early 2000s the several groups each independently reported the optical microscope measurements of the ballistic switching by picosecond pulse magnetic field. \cite{gerrits_ultrafast_2002, tudosa_ultimate_2004, schumacher_quasiballistic_2003, hiebert_fully_2003, hiebert_spatially_2003}. Then the mechanism of a ballistic switching was analyzed in terms of the nonlinear dynamics concepts such as a fixed point, attractors, and saddle point.\cite{sun_fast_2005, xiao_minimal_2006, xiao_effect_2007} Especially the minimal field required for a ballistic switching was investigated by comparing the so-called Stoner-Wohlfarth (SW) type. \cite{sun_fast_2005, xiao_minimal_2006}  The damping constant dependence of the minimal switching field was also studied. \cite{sun_fast_2005} The characteristics of the parameters of a pulse magnetic field, i.e., magnitude, direction, and rise/fall time on the mechanism of a ballistic switching had been also studied by the simulations and experiments.\cite{xiao_effect_2007, schumacher_ballistic_nodate, horley_magnetization_2010, kikuchi_quasi-ballistic_2014}. 

As described above, in 2000s and 2010s a ballistic switching technique had received much attention for the fast magnetization reversal with ringing suppression by fine-tuning the magnetic pulse parameters. Due to the recent advance of an ultra-fast measurement\cite{neeraj_inertial_2021} the studies of a ballistic switching have attracted much attention again. Last year the in-plane magnetization switching dynamics as functions of the pulse magnetic field duration and amplitude was calculated and analyzed by using the conventional Landau-Lifshitz-Gilbert (LLG) equation and its inertial form, the so-called iLLG equation\cite{neeraj_magnetization_2022}. The solutions of both equations were compared in terms of the switching characteristics, speed and energy density analysis. Both equations return qualitatively similar switching dynamics. However the extensive material parameter dependences of a ballistic switching region have not yet been sufficiently explored. Therefore it is worth clearing the extensive material parameter dependences of the ballistic switching of an in-plane magnetized nanomagnet.

In this paper, we study the ballistic switching of an in-plane magnetized nanomagnet with systematically varying the material parameters by using the macrospin simulations. The results show that the pulse width required for the ballistic switching can be reduced by increasing the magnetic anisotropy constant or by decreasing the saturation magnetization. There exists an optimal value of the Gilbert damping constant that minimizes the pulse width required for ballistic switching. The simulation results are intuitively explained by analyzing the switching trajectory on the energy density contour.

\section{Model and Method}
In this section we show the theoretical model, the numerical simulation method, and the analysis using the trajectory in the limit of $\alpha\to 0$. The macrospin model of the in-plane magnetized noanomagnet and the equations we solve to simulate the magnetization dynamics are given in Sec. \ref{sec:macro}. In Sec. \ref{sec:analysis} we show that the switching conditions can be analyzed by using the trajectory on the energy density contour in the limit of $\alpha\to 0$ if the $\alpha \ll 1$.

\subsection{Macrospin Model Simulation}
\label{sec:macro}
\begin{figure}[t]
    \begin{center}
        \includegraphics[width=0.95\columnwidth,clip]{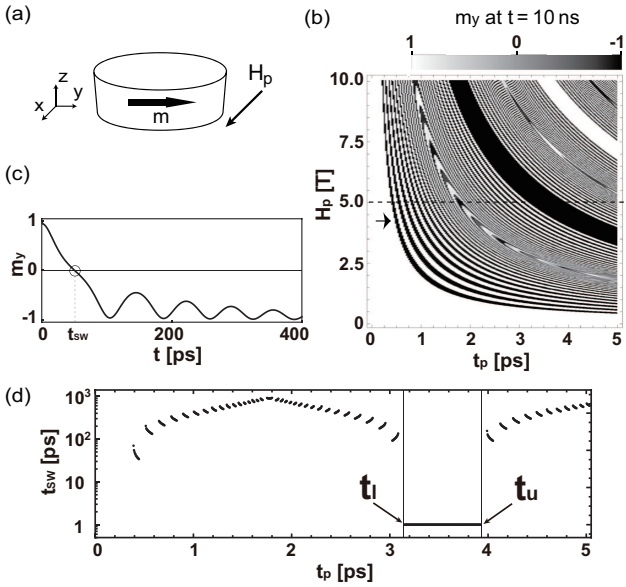}
        \caption{(a) Schematic illustration of the in-plane magnetized nanomagnet. The pulse field, $H_{\rm p}$, is applied along the $x$-direction. The initial direction of the magnetization is in the positive $y$-direction.
            (b) Gray scale map of $m_{\rm y}$ at $t=10$ ns as a function of the pulse field width, $t_{\rm p}$, and $H_{\rm p}$. The black and white regions represent the success and failure of switching. The parameters are $\mu_{0} M_{\rm s}$ = 0.92 T, $\mu_{0} H_{\rm K}$ = 0.1 T, and $\alpha$ = 0.023.
            (c) Typical example of the time evolution of $m_{\rm y}$ when the magnetization switches ($H_{\rm p}$ = 5 T and $t_{\rm p}$ = 0.4 ps). The switching time, $t_{\rm SW}$, is defined as the time when $m_{\rm y}$ changes the sign.
            (d) $t_{\rm p}$ dependence of $t_{\rm SW}$ along the dashed horizontal line at $H_{\rm p}=5$ T shown in Fig. \ref{fig1}(b). $t_{\rm l}$ and $t_{\rm u}$ are 3.15 ps and 3.93 ps, respectively. $t_{\rm SW}$ at $t_{\rm l} \le t_{\rm p} \le t_{\rm u}$ is 1.7 ps. 
        }
        \label{fig1}
    \end{center}
\end{figure}

Figure \ref{fig1}(a) shows the schematic illustration of the in-plane magnetized nanomagnet. The pulsed magnetic field, $\bm{H}_{\rm p}$, is applied along the $x$-direction. The unit vector $\bm{m}$ = $(m_{\rm x}, m_{\rm y}, m_{\rm z})$ indicates the direction of the magnetization. The size of the nanomagnet is assumed to be so small that the dynamics of $\bm{m}$ can be described by the macrospin LLG equation
\begin{align}
    \label{eq:llglab}
    \frac{d\bm{m}}{dt}
    =
    -\gamma\bm{m}\times
    \left({\bm H}_{\rm eff}
    -\frac{\alpha}{\gamma}\frac{d{\bm m}}{dt}
    \right),
\end{align}
where $t$ is time, $\gamma$ is the gyromagnetic ratio,
$\alpha$ is the Gilbert damping constant. The effective field,
$
    \bm{H}_{\rm eff}
    = \bm{H}_{\rm p}
    + \bm{H}_{\rm K}
    + \bm{H}_{\rm d},
$
comprises the pulse field, $\bm{H}_{\rm p}$, the anisotropy field, $\bm{H}_{\rm K}$, and the demagnetizing field, $\bm{H}_{\rm d}$.
The anisotropy field and the demagnetizing field are defined as
\begin{equation}
    \bm{H}_{\rm K}
    =
    \left[2K/(\mu_0 M_{\rm s})\right]m_{y}\bm{e}_y,
\end{equation}
and
\begin{equation}
    \bm{H}_{\rm d}
    =
    \mu_0 M_{\rm s}m_{z}\bm{e}_z,
\end{equation}
respectively, where $K$ is the uniaxial anisotropy constant, $\mu_0$ is the magnetic
permeability of vacuum, $M_{\rm s}$ is the saturation
magnetization, and $\bm{e}_{\rm j}$ is the unit vector along the $j$-axis ($j$ = $x$, $y$, $z$).

The switching dynamics are calculated by numerically solving the LLG equation. The initial ($t=0$) direction is set as $m_{\rm y}$ = 1. The rectangular shaped pulse magnetic field with duration of $t_{\rm p}$ is applied at $t=$0. The time evolution of magnetization dynamics are calculated for 10 ns. Success or failure of switching is determined by whether $m_{\rm y} < -0.5$ at $t=$ 10 ns.

Figure \ref{fig1}(b) shows the gray scale plot of $m_{y}$ at $t=$ 10 ns on the $t_{\rm p}$-$H_{\rm p}$ plane.  Following Ref. \citenum{neeraj_magnetization_2022} the parameters are assumed to be $\mu_{0} M_{\rm s}$ = 0.92 T, $K=2.3$ kJ/m$^3$, i.e. $\mu_{0} H_{\rm K}$ = 0.1 T, and $\alpha$ = 0.023. The black and white regions represent the success and failure of switching, respectively. The wide black region at upper right of Fig. \ref{fig1}(b) represents the ballistic switching region (BSR). A typical example of the time evolution of $m_{\rm y}$ when the magnetization switches is shown in Fig. \ref{fig1}(c). The switching time, $t_{\rm SW}$, is defined as the time when $m_{\rm y}$ changes the sign. Figure \ref{fig1}(d) shows the $t_{\rm p}$ dependence of $t_{\rm SW}$ along the horizontal line shown in Fig. \ref{fig1}(b), i.e. at $H_{\rm p}=5$ T. The BSR indicated by shade appears between $t_{\rm l}=3.15$ ps and $t_{\rm u} = 3.93$ ps, where $t_{\rm SW} = 1.7$ ps independent of $t_{\rm p}$. The lower and upper boundary of the BSR are represented by $t_{\rm l}$ and $t_{\rm u}$, respectively. We investigate the material parameter dependence of $t_{\rm l}$ and $t_{\rm u}$ with keeping $H_{\rm p} = 5$ T.

\subsection{Analysis of the Switching Conditions for $\alpha \ll 1$}
\label{sec:analysis}
\begin{figure}[t]
    \begin{center}
        \includegraphics[width=0.81\columnwidth,clip]{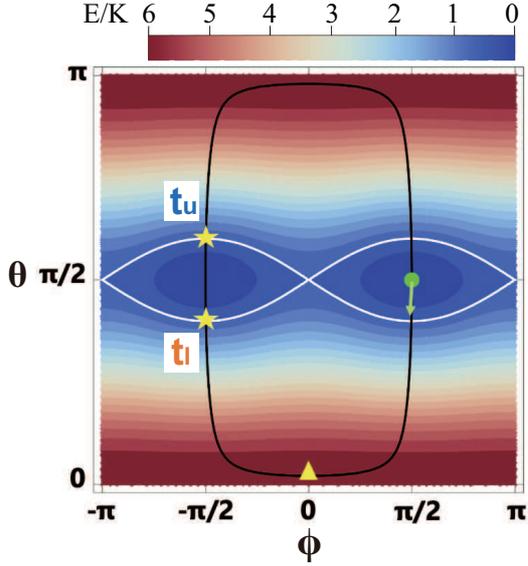}
        \caption{
            (Color online) Color plot of the energy density contour given by Eq. \eqref{eq:energy_density}. $\theta$ and $\phi$ are the polar and azimuthal angles of the magnetization, respectively.  The material parameters, $M_{\rm s}$ and $K$ are same as in Fig. \ref{fig1}. The separatrix given by Eq. \eqref{eq:separatrix} is indicated by the white curve. The initial direction of $\bm{m}$ is indicated by the green dot at $(\theta, \phi) = (\pi/2, \pi/2)$. The black curve represents the trajectory of the magnetization under the field of $H_{\rm p} = 5$ T in the limit of $\alpha \to 0$, which is given by Eq. \eqref{eq:trajectory}. The yellow stars indicate the intersection points of the separatrix and the trajectory, which correspond to $t_{\rm p}$ =  $t_{\rm l}$ and $t_{\rm u}$. If the pulse is turned off at $t_{\rm l} \le t \le t_{\rm u}$, the magnetization switches ballistically. The yellow triangle indicates the turning point of the trajectory of the magnetization near $m_{z}=1$, at which $\phi=0$.
        }
        \label{fig2}
    \end{center}
\end{figure}

If the Gilbert damping constant is much smaller than unity the approximate value of $t_{\rm l}$ and $t_{\rm u}$ can be obtained without performing macrospin simulations. In the limit of $\alpha \rightarrow$ 0, the trajectory is represented by the energy contour because the energy is conserved during the motion of $\bm{m}$.  The energy density, $E$, of the nanomagnet is defined as\cite{brown_thermal_1963}
\begin{equation}
    \label{eq:energy_density}
    E= \frac{1}{2}\mu_{0}M_{\rm s}^{2}\cos^2\theta + K(1 - \sin^2\theta\sin^2\phi),
\end{equation}
where $\theta$ and $\phi$ are the polar and azimuthal angles of the magnetization, respectively. The color plot of the energy density contour is shown in Fig. \ref{fig2}. The separatrix representing the energy contour with $E$ = $K$ is indicated by the white curve, which is expresses as
\begin{equation}
    \label{eq:separatrix}
    \frac{1}{2}\mu_{0}M_{\rm s}^{2}\cos^2\theta - K\sin^2\theta\sin^2\phi = 0.
\end{equation}
The green dot indicates the initial direction of $\bm{m}$ at $t=0$. The black curve represents the trajectory of $\bm{m}$ under the pulse field of $H_{\rm p}$ in the limit of $\alpha \to 0$. Under the pulse field the energy density is given by
\begin{align}
    \label{eq:energy_density_hp}
    E  = \frac{1}{2}\mu_{0}M_{\rm s}^{2}\cos^2\theta & + K(1 - \sin^2\theta\sin^2\phi)\notag          \\
                                                     & - \mu_{0}M_{\rm s}H_{\rm p}\sin\theta\cos\phi.
\end{align}
Since the energy density of the initial direction, $\theta=\phi=\pi/2$, is $E=0$, the trajectory under the pulse field is expressed as
\begin{align}
    \label{eq:trajectory}
    \frac{1}{2}\mu_{0}M_{\rm s}^{2}\cos^2\theta & + K(1 - \sin^2\theta\sin^2\phi)\notag                   \\
                                                & - \mu_{0}M_{\rm s}H_{\rm p}\sin\theta\cos\phi =0.
\end{align}

The yellow stars indicate the points where the trajectory crosses the separatrix surrounding the equilibrium point at $\phi=-\pi/2$. The upper and lower points indicates the direction of $\bm{m}$ at the end of the pulse with $t_{\rm p}=t_{\rm u}$ and $t_{\rm l}$, respectively. The corresponding angles $(\theta_{\rm l}, \phi_{\rm l})$ and $(\theta_{\rm u}, \phi_{\rm u})$ can be obtained by solving Eqs. \eqref{eq:separatrix} and \eqref{eq:trajectory} simultaneously. If $t_{\rm l} \le t_{\rm p} \le t_{\rm u}$, the magnetization relaxes to the equilibrium direction at $(\theta, \phi) = (\pi/2, -\pi/2)$ after the pulse to complete the switching.  We can obtain the approximate expressions of $t_{\rm l}$ and $t_{\rm u}$ as follows. Assuming that the pulse field is much larger than the other fields, the angular velocity of the precession, $\omega$, is approximated as $\gamma H_{\rm p}/(1+\alpha^2)$, and $t_{\rm l}$ and $t_{\rm u}$ are analytically obtained as
\begin{equation}
    \label{eq:t_lower}
    t_{\rm l} = \frac{\pi -2\theta_{\rm turn}}{\omega}-\frac{1}{2}\frac{\Delta\theta}{\omega},
\end{equation}
and
\begin{equation}
    \label{eq:t_upper}
    t_{\rm u} = \frac{\pi -2\theta_{\rm turn}}{\omega}+\frac{1}{2}\frac{\Delta\theta}{\omega},
\end{equation}
where $\Delta\theta = \theta_{\rm u} - \theta_{\rm l}$, and $\theta_{\rm turn}$ is the polar angle at the turning point ($\phi=0$) indicated by the yellow triangle.

\section{Results and Discussion}
In this section we discuss the dependence of the BSR on the material parameters by analyzing the numerical simulation results and Eqs. \eqref{eq:t_lower} and \eqref{eq:t_upper}. The results for the variation of the magnetic anisotropy constant, $K$, saturation magnetization, $M_{s}$, and the Gilbert damping constant, $\alpha$, will be given in Secs. \ref{sec:kdep}, \ref{sec:mdep}, and \ref{sec:adep}, respectively.

\subsection{Anisotropy Constant Dependence of the BSR}
\label{sec:kdep}
\begin{figure}[t]
    \begin{center}
        \includegraphics[width=0.85\columnwidth,clip]{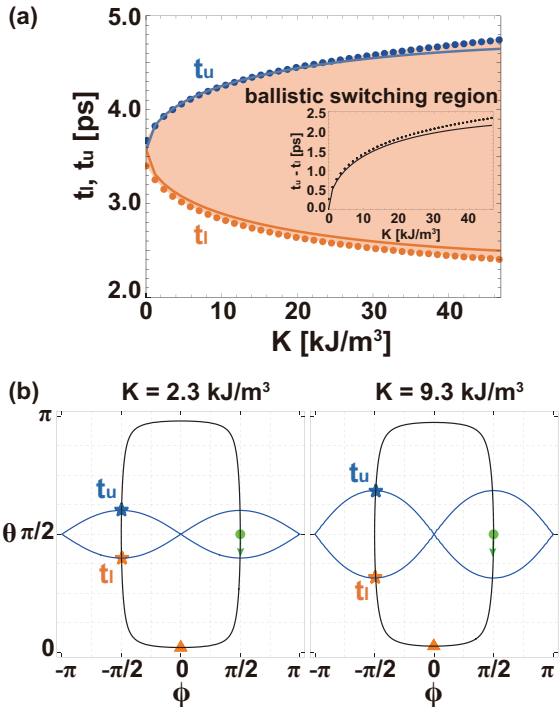}
        \caption{
            (Color online) (a) Anisotropy constant, $K$, dependence of the BSR (orange shade). Simulation results of $t_{\rm l}$ and $t_{\rm u}$ are plotted by the orange and blue dots, respectively. The analytical results are indicated by the solid curves with the same color. The parameters are $H_{\rm p} = 5$ T, $\mu_{0} M_{\rm s} = 0.92$ T, and $\alpha$ = 0.023. In the inset the simulation and analytical results of the width of the BSR, $t_{\rm u}$ - $t_{\rm l}$, are plotted by the dots and the solid curve, respectively.
            (b) Typical examples of the trajectory of the magnetization (black curve) and the separatrix (blue curve). The left and right panels show the results for $K$ = 2.3 kJ/m$^3$ and $K$ = 9.3 kJ/m$^3$, respectively. The orange and blue stars indicate the direction at $t = t_{\rm l}$ and $t_{\rm u}$, respectively. The green dots indicate the initial direction of $\bm{m}$.
        }
        \label{fig3}
    \end{center}
\end{figure}

Figure \ref{fig3}(a) shows the anisotropy constant, $K$, dependence of the BSR. The parameters are $H_{\rm p} = 5$ T, $\mu_{0} M_{\rm s} = 0.92$ T, and $\alpha$ = 0.023. The simulation results of $t_{\rm l}$ and $t_{\rm u}$ are indicated by the orange and blue dots, respectively. The analytical approximations of $t_{\rm l}$ and $t_{\rm u}$ obtained by solving Eqs. \eqref{eq:separatrix},\eqref{eq:trajectory},\eqref{eq:t_lower}, and \eqref{eq:t_upper} are represented by the orange and blue curves, respectively. The simulation and analytical results agree well with each other because the Gilbert damping constant is as small as 0.023. As shown in Fig. \ref{fig3}(a), $t_{\rm l}$ is a monotonically decreasing function of  $K$ while $t_{\rm u}$ is a monotonically increasing function of $K$. As a result the width of the BSR, $t_{\rm u}$ - $t_{\rm l}$, is a monotonically increasing function of $K$ as shown in the inset of Fig. \ref{fig3}(a).

In the  left panel of Fig. \ref{fig3}(b) the separatrix and the trajectory with $\alpha=0$ for $K= 2.3$ kJ/m$^{3}$ are shown by the blue and black curves, respectively. The same plot for $K= 9.3$ kJ/m$^{3}$ is shown in the right panel. As shown in these panels, the increase of $K$ does not change the trajectory much.  However, the increase of $K$ changes the separatrix significantly through the second term of Eq. \eqref{eq:separatrix}. Assuming that the angular velocity of the precession is almost constant, the spread of the area surrounded by the separatrix results in the spread of the time difference between $t_{\rm l}$ and $t_{\rm u}$. As a result the BSR is spread by the increase of $K$ as shown in Fig. \ref{fig3}(a)

\subsection{Saturation Magnetization Dependence of the BSR}
\label{sec:mdep}
\begin{figure}[t]
    \begin{center}
        \includegraphics[width=0.85\columnwidth,clip]{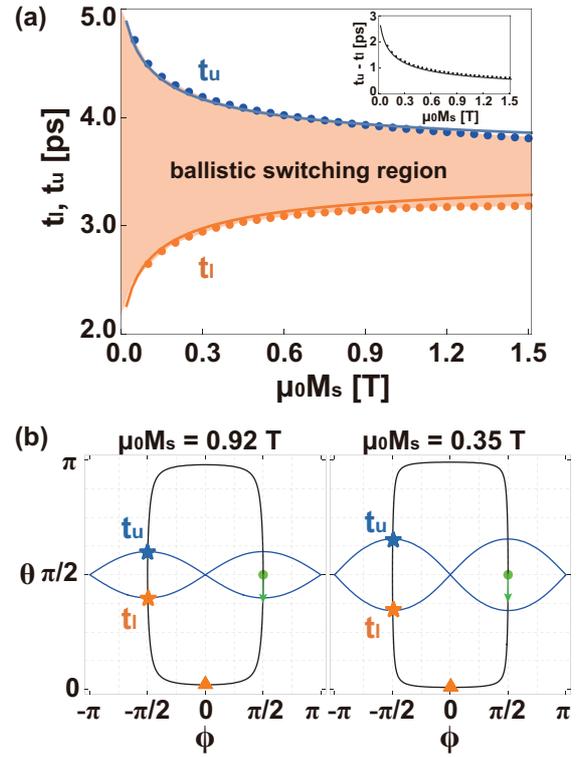}
        \caption{
             (Color online) (a) Saturation magnetization dependence of the BSR. The horizontal axis represents the saturation magnetization in unit of T, i.e $\mu_{0} M_{s}$.  The parameters are $H_{\rm p} = 5$ T, $K = 2.3$ kJ/m$^{3}$, and $\alpha$ = 0.023. The symbols are the same as in Fig. \ref{fig3} (a).
            (b) Typical examples of the trajectory of the magnetization (black curve) and the separatrix (blue curve). The right and left panels show the results for $\mu_{0} M_{s} =$ 0.35 T and 0.92 T, respectively. The symbols are the same as in Fig. \ref{fig3} (b).
        }
        \label{fig4}
    \end{center}
\end{figure}

Figure \ref{fig4}(a) shows the saturation magnetization dependence of the BSR obtained by the numerical simulation and the analytical approximations. The horizontal axis represents the saturation magnetization in unit of T, i.e $\mu_{0} M_{s}$. The parameters are $H_{\rm p} = 5$ T, $K = 2.3$ kJ/m$^{3}$, and $\alpha$ = 0.023. The symbols are the same as in Fig. \ref{fig3}(a). The lower boundary of the BSR, $t_{\rm l}$, increases as the $\mu_{0} M_{\rm s}$ increases while the upper boundary of the BSR, $t_{\rm u}$, decreases with increase of $\mu_{0} M_{\rm s}$. Therefore, the faster switching is available for smaller $M_{s}$. The $\mu_{0} M_{\rm s}$ dependence of the BSR ($t_{\rm u}$ - $t_{\rm l}$) is also shown in the inset of Fig. \ref{fig4}(a). The BSR decreases with increase of $\mu_{0} M_{s}$. In other words, the wider BSR is obtained for smaller $M_{s}$.

In the right panel of Fig. \ref{fig4}(b) the separatrix and the trajectory with $\alpha=0$ for $\mu_{0}M_{s}= 0.35$ T are shown by the blue and black curves, respectively. The same plot for $\mu_{0}M_{s}= 0.92$ T is shown in the left panel. As shown in these panels, the increase of $M_{s}$ does not change the trajectory much but decrease the separatrix significantly through the first term of Eq. \eqref{eq:separatrix}. Assuming that the angular velocity of the precession is almost constant, the reduction of the area surrounded by the separatrix results in the reduction of the time difference between $t_{\rm l}$ and $t_{\rm u}$. As a result the BSR decreases with increase of $M_{s}$ as shown in Fig. \ref{fig4}(a)

\subsection{Gilbert Damping Constant Dependence of the BSR}
\begin{figure}[t]
    \begin{center}
        \includegraphics[width=0.7\columnwidth,clip]{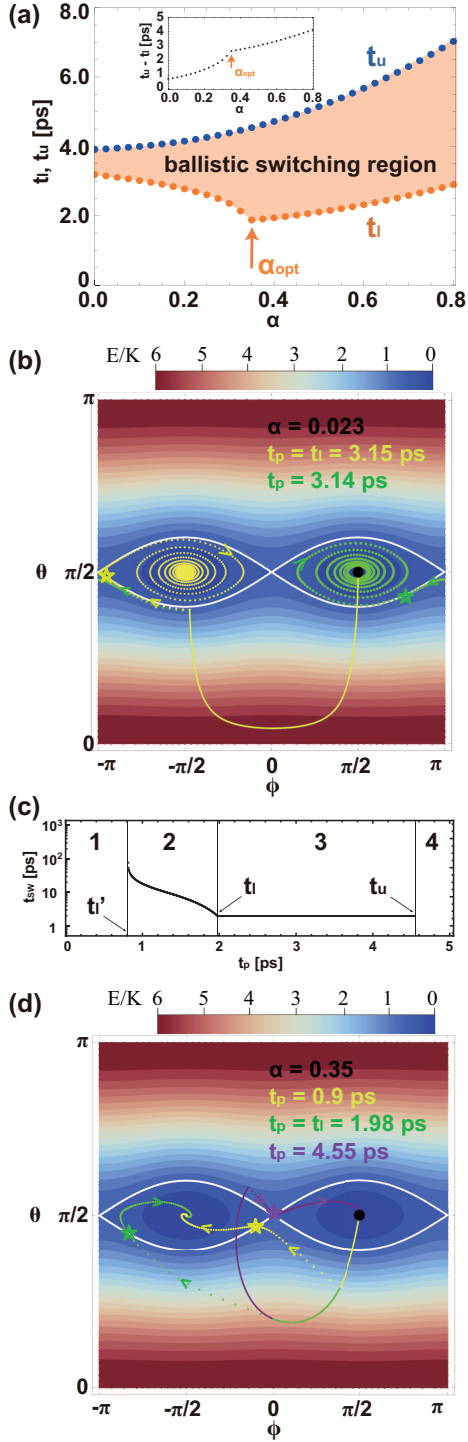}
        \caption{(Color online) (a) Simulation results of the $t_{\rm l}$ and $t_{\rm u}$  as a function of $\alpha$. The parameters are $H_{\rm p}$ = 5.0 T, $K = 2.3$ kJ/m$^{3}$, and $\mu_{0} M_{\rm s}$ = 0.92 T. The symbols are the same as in Fig. \ref{fig3} (a).
            (b) The trajectories at $\alpha=0.023$ with $t_{\rm p} = 3.15$ ps (yellow) and 3.14 ps (green) on the energy density contour. The trajectory during the pulse is represented by the solid curve. The trajectory after the pulse is represented by the dots. The white curve shows the separatrix.The direction of the trajectory is indicated by the arrow. The star indicates the intersection point of the trajectory and the separatrix. The initial direction is indicated by the black dot.
            (c) $t_{\rm p}$ dependence of $t_{\rm SW}$ at $\alpha=\alpha_{\rm opt}$. All parameters except $\alpha$ are the same as in Fig. \ref{fig1} (d). $t_{\rm l}'$, $t_{\rm l}$, and $t_{\rm u}$ are 0.82 ps, 1.98 ps, and 4.54 ps, respectively. $t_{\rm SW}$ at $t_{\rm l} \le t_{\rm p} \le t_{\rm u}$ is 1.98 ps.
            (d) The trajectories at $\alpha=\alpha_{\rm opt}$ with $t_{\rm p} = 0.9$ ps (yellow), 1.98 ps (green), and 4.55 ps (purple) on the energy density contour. The symbols are the same as in Fig. \ref{fig5} (b). 
        }
        \label{fig5}
    \end{center}
\end{figure}
\label{sec:adep}
Figure \ref{fig5}(a) shows the simulation results of the Gilbert damping constant, $\alpha$, dependence of the BSR. The width of the BSR is shown in the inset. The symbols are the same as in Fig. \ref{fig3}(a). The approximate values obtained by Eqs. \eqref{eq:t_lower} and \eqref{eq:t_upper} are not shown because the $\alpha$ is not limited to $\alpha \ll 1$. The parameters are $H_{\rm p} = 5$ T, $K= 2.3$ kJ/m$^{3}$, and $\mu_{0} M_{\rm s} = 0.92$ T. There exists an optimal value of $\alpha$ that minimizes $t_{\rm l}$. The optimum value in Fig. \ref{fig5} (a) is $\alpha_{\rm opt}$ = 0.35.

To understand the mechanism for minimization of $t_{\rm l}$ at a certain value of $\alpha$ one need to consider two different effects of $\alpha$ on the magnetization dynamics.
One effect is the decrease of the precession angular velocity with increase of $\alpha$. The precession angular velocity around the effective field of $H_{\rm eff}$ is given by $(\gamma H_{\rm eff})/(1+\alpha^{2})$, which decreases with increase of $\alpha$. This effect causes the increase of $t_{\rm l}$ and $t_{\rm u}$.

The other effect is the increase of the energy dissipation rate with increase of $\alpha$. Let us consider the trajectory in the cases of small damping  ($\alpha 
= 0.023$) and large damping ($\alpha = \alpha_{\rm opt}$). In Fig. \ref{fig5} (b) the typical examples of the trajectory for the small damping are shown by the yellow and green curves and dots on the energy density contour. The pulse widths are $t_{\rm p}=t_{\rm l}(=3.15$ ps) and $3.14$ ps. The trajectories during the pulse are represented by the solid curves and the trajectories after the pulse are represented by the dots. The white curve shows the separatrix and the black dot indicates the initial direction. The yellow and green stars indicate the points where the trajectories cross the separatrix surrounding the target and initial states, respectively. The arrows indicate the direction of the movement of the magnetization. For the small damping, even very close to the separatrix around the target state at the end of the pulse, the magnetization flows to the sepatrarix around the initial state and relax to the initial state after many precessions with the slow energy dissipation.

Figure \ref{fig5} (c) shows the $t_{\rm p}$ dependence of $t_{\rm SW}$ at the large damping ($\alpha = \alpha_{\rm opt}$). All parameters except $\alpha$ are the same as in Fig. \ref{fig1} (d). $t_{\rm l}'$, $t_{\rm l}$, and $t_{\rm u}$ are 0.82 ps, 1.98 ps, and 4.54 ps, respectively. $t_{\rm l}'$ is the time when for the large damping the magnetization goes across the effective separatrix around the initial state during the pulse duration.
In Fig. \ref{fig5} (d) the typical examples of the trajectory for the large damping are shown by the yellow ($t_{\rm p}=0.9$ ps), green ($t_{\rm p}=t_{\rm l}=1.98$ ps), and purple ($t_{\rm p}=4.55$ ps) curves and dots on the energy density contour. The symbols are the same as in Fig. \ref{fig5} (b). 
In the region 1 ($t_{\rm p} < t_{\rm l}'$) of Fig. \ref{fig5} (c) the magnetization is located on the inside of the effective separatrix at the end of the pulse and return to the initial state. The trajectory is not shown in Fig. \ref{fig5} (d). In the region 2 ($t_{\rm l}' \le t_{\rm p} < t_{\rm l}$) of Fig. \ref{fig5} (c) after the pulse is removed the magnetization move toward the target state under the effective field of $H_{\rm eff}$ and goes across the separatrix. Then the magnetization finishes the switching. The typical example of the trajectory is shown by the yellow curve and dots in Fig. \ref{fig5} (d). In the region 3 ($t_{\rm l} \le t_{\rm p} \le t_{\rm u}$) after the pulse is removed the magnetization ballistically switches. The typical example of the trajectory is shown by the green curve and dots in Fig. \ref{fig5} (d). As explained in Sec. \ref{sec:macro}, in this study the $t_{\rm SW}$ is defined as the time when $m_{\rm y}$ changes the sign, in other words, the magnetization reaches the turning point ($\phi=0$) on the energy density contour. Therefore the $t_{\rm l}$ for the large damping is equal to the ballistic $t_{\rm SW}$, i.e. the time when under the pulse field the magnetization reaches the turning point. The ballistic $t_{\rm SW}$ at the region 3 is 1.98 ps. 

As described above, for the large damping the magnetization can relax to the target state even at greater distances from the separatrix by moving under $H_{\rm eff}$ with the fast energy dissipation. The effect can be regarded as the effective spread of the separatrix and results in the decrease of $t_{\rm l}$ and the increase of $t_{\rm u}$ with increase of $\alpha$. Therefore, there exists the optimal value of $\alpha$ that minimizes $t_{\rm l}$ while $t_{\rm u}$ monotonically increases with increase of $\alpha$ as shown in Fig. \ref{fig5}(a). In the region 4 ($t_{\rm p} > t_{\rm u}$) after the pulse is removed the magnetization moves toward the separatrix around the initial state under $H_{\rm eff}$ and relaxes to the initial state. We find that the BSR for the large damping can be explained by the anisotropic spread of the effective separatrix with increasing $\alpha$, which is fundamentally due to the breaking of the spatial inversion symmetry of the spin dynamics. The broken symmetry of the spatial inversion of the spin dynamics for the large damping can be easily confirmed by comparing Fig. \ref{fig5} (c) with Fig. \ref{fig1} (d). 

\section{Summary}
In summary, we study the material parameter dependence of the ballistic switching region of the in-plane magnetized nanomagnets based on the macrospin model. The results show that the pulse width required for the ballistic switching can be reduced by increasing the magnetic anisotropy constant or by decreasing the saturation magnetization. The results also revealed that there exists an optimal value of the Gilbert damping constant that minimizes the pulse width required for the ballistic switching. The simulation results are explained by analyzing the trajectories on the energy contour. The results are useful for further development of the high-speed information processing using the ballistic switching of magnetization.

\begin{acknowledgments}
   This work is partially supported by JSPS KAKENHI Grant Number JP20K05313.
\end{acknowledgments}

\bibliographystyle{jpsj}
\bibliography{references}

\begin{thebibliography}{10}

\bibitem{he_theoretical_1996}
L.~He and W.~D. Doyle: J. Appl. Phys. {\bfseries 79} (1996) 6489.

\bibitem{sun_fast_2005}
Z.~Z. Sun and X.~R. Wang: Phys. Rev. B {\bfseries 71} (2005) 174430.

\bibitem{xiao_minimal_2006}
D.~Xiao, M.~Tsoi, and Q.~Niu: Journal of Applied Physics {\bfseries 99} (2006)
  013903.

\bibitem{nozaki_numerical_2006}
Y.~Nozaki and K.~Matsuyama: Jpn. J. Appl. Phys. {\bfseries 45} (2006) L758.

\bibitem{nozaki_numerical_2006-1}
Y.~Nozaki and K.~Matsuyama: Journal of Applied Physics {\bfseries 100} (2006)
  053911.

\bibitem{xiao_effect_2007}
Q.~F. Xiao, B.~C. Choi, J.~Rudge, Y.~K. Hong, and G.~Donohoe: Journal of
  Applied Physics {\bfseries 101} (2007) 024306.

\bibitem{horley_magnetization_2010}
P.~P. Horley, V.~R. Vieira, P.~M. Gorley, V.~K. Dugaev, J.~Berakdar, and
  J.~Barna^^c5^^9b: Journal of Magnetism and Magnetic Materials {\bfseries 322}
  (2010) 1373.

\bibitem{bazaliy_analytic_2011}
Y.~B. Bazaliy: Journal of Applied Physics {\bfseries 110} (2011) 063920.

\bibitem{gerrits_ultrafast_2002}
T.~Gerrits, H.~A.~M. van~den Berg, J.~Hohlfeld, L.~B^^c3^^a4r, and T.~Rasing:
  Nature {\bfseries 418} (2002) 509.

\bibitem{tudosa_ultimate_2004}
I.~Tudosa, C.~Stamm, A.~B. Kashuba, F.~King, H.~C. Siegmann, J.~St^^c3^^b6hr,
  G.~Ju, B.~Lu, and D.~Weller: Nature {\bfseries 428} (2004) 831.

\bibitem{schumacher_quasiballistic_2003}
H.~W. Schumacher, C.~Chappert, R.~C. Sousa, P.~P. Freitas, and J.~Miltat: Phys.
  Rev. Lett. {\bfseries 90} (2003) 017204.

\bibitem{hiebert_fully_2003}
W.~K. Hiebert, L.~Lagae, J.~Das, J.~Bekaert, R.~Wirix-Speetjens, and
  J.~De~Boeck: Journal of Applied Physics {\bfseries 93} (2003) 6906.

\bibitem{hiebert_spatially_2003}
W.~K. Hiebert, L.~Lagae, and J.~De~Boeck: Phys. Rev. B {\bfseries 68} (2003)
  020402.

\bibitem{schumacher_ballistic_nodate}
H.~W. Schumacher: Appl. Phys. Lett. {\bfseries 87} (2005) 042504.

\bibitem{kikuchi_quasi-ballistic_2014}
N.~Kikuchi, Y.~Suyama, S.~Okamoto, O.~Kitakami, and T.~Shimatsu: Appl. Phys.
  Lett. {\bfseries 104} (2014) 112409.

\bibitem{neeraj_magnetization_2022}
K.~Neeraj, M.~Pancaldi, V.~Scalera, S.~Perna, M.~d'Aquino, C.~Serpico, and
  S.~Bonetti: Phys. Rev. B {\bfseries 105} (2022) 054415.

\bibitem{neeraj_inertial_2021}
K.~Neeraj, N.~Awari, S.~Kovalev, D.~Polley, N.~Zhou~Hagstr^^c3^^b6m, S.~S.
  P.~K. Arekapudi, A.~Semisalova, K.~Lenz, B.~Green, J.-C. Deinert, I.~Ilyakov,
  M.~Chen, M.~Bawatna, V.~Scalera, M.~d’Aquino, C.~Serpico, O.~Hellwig, J.-E.
  Wegrowe, M.~Gensch, and S.~Bonetti: Nat. Phys. {\bfseries 17} (2021) 245.

\bibitem{brown_thermal_1963}
W.~F. Brown: Phys. Rev. {\bfseries 130} (1963) 1677.

\end{thebibliography}

\end{document}